\documentclass[eqsecnum,nofootinbib,floats,preprint,aps,floatfix,titlepage,tightenlines]{revtex4} 
\input epsf 
\input graphicx

\def\Tr{{\rm Tr\,}}

\makeatletter

\def\vereq#1#2{\lower3pt\vbox{\baselineskip1.5pt \lineskip1.5pt
\ialign{$\m@th#1\hfill##\hfil$\crcr#2\crcr\sim\crcr}}}
\def\lesssim{\mathrel{\mathpalette\vereq<}}
\@addtoreset{equation}{section} \makeatother

\begin{document}

\title{Instabilities of chromodyons in SO(5) gauge theory}

\author{Huidong Guo}
\email{hg90@columbia.edu}
\author{Erick J. Weinberg}
\email{ejw@phys.columbia.edu}
\affiliation{Physics Department, Columbia University, New York, New York 10027
\vskip 0.5in}

\begin{abstract}

Attempts to construct chromodyons --- objects with both magnetic
charge and non-Abelian electric charge --- in the context of
spontaneously broken gauge theories have been thwarted in the past by
topological obstructions to globally defining the unbroken non-Abelian
``color'' subgroup.  In this paper we consider the possibility of
chromodyons in a theory with SO(5) broken to ${\rm SU(2)}\times{\rm
U(1)}$, where the topological obstructions are absent.  We start by
constructing a monopole with only magnetic charge.  By exciting a
global gauge zero mode about this monopole, we obtain a chromodyonic
configuration that is an approximate solution of the field equations.
We then numerically simulate the time evolution of this initial state,
to see if it settles down in a stationary solution.  Instead, we find
that chromo-electric charge is continually radiated away, with every
indication that this process will continue until this charge has been
completely lost.  We argue that this presents strong evidence against the
existence of stable chromodyons.

\end{abstract}

\preprint{CU-TP-1183}

\maketitle

\section{Introduction}

The magnetic monopoles that arise in spontaneously broken gauge
theories can easily be generalized to dyons that have a U(1) electric
charge in addition to their magnetic charge.  It then is natural to
ask whether, in cases where the unbroken symmetry is non-Abelian, it
is possible to have monopoles carrying non-Abelian electric charge.
Such objects, referred to as chromodyons, were first considered in the
context of an SU(5) grand unified theory.  After attempts to construct
such chromodyons failed~\cite{Abouelsaood:1982dz}, it was shown that
the non-Abelian magnetic charge of the SU(5) monopole creates a
topological obstruction to the existence of non-Abelian ``color''
electric
charge~\cite{Nelson:1983bu,Balachandran:1982gt,Balachandran:1983xz,%
Balachandran:1983fg,Horvathy:1985bp,Horvathy:1984yg},
and the issue was abandoned for a number of years.  Since then,
however, it has been realized that, with other choices of gauge group,
there can be monopoles with purely Abelian magnetic charge, even
though the unbroken gauge group is non-Abelian~\cite{Weinberg:1982jh}.
Because there is then no topological barrier to a color electric
charge, it is natural to revisit the subject and consider whether
chromodyons can exist.

Let us first recall how ordinary U(1) dyons arise.  When there is an
unbroken U(1) symmetry, any soliton with a nonvanishing charged field
has a zero mode corresponding to a shift in the phases of the complex
charged fields.  Exciting this mode in a time-dependent fashion
produces a U(1) charge.  If the U(1) symmetry is gauged, it may be
possible to gauge transform away the time-dependence of the phase, but
the gauge-invariant electric charge remains.  The simplest example of
this occurs with SU(2) broken to U(1), where the Julia-Zee
dyon~\cite{Julia:1975ff} arises from rotation of the phase of the
massive vector boson fields in the core of the 't~Hooft-Polyakov
monopole~\cite{'t Hooft:1974qc,Polyakov:1974ek}.

Similarly, if there is an unbroken non-Abelian symmetry, excitation of
the gauge orientation zero modes of a soliton gives rise to a
non-Abelian electric charge.  In the theory with SU(5) broken to ${\rm
SU(3)}_{\rm color}\times{\rm U(1)}_{\rm EM}$, the unit monopoles have
nontrivial fields that are not invariant under the unbroken SU(3).
Hence, one would expect to be able to generate dyons that were charged
under the color SU(3) (hence the term ``chromodyon'') by exciting the
resulting global gauge zero modes.  However,
Abouelsaood~\cite{Abouelsaood:1982dz} found that, because the gauge
potential has a $1/r$ tail in the unbroken subgroup, some of the
expected zero modes are non-normalizable, and the proposed
construction does not go through.  A deeper explanation for this was
given by Nelson and Manohar~\cite{Nelson:1983bu}, and by Balachandran
et
al.~\cite{Balachandran:1982gt,Balachandran:1983xz,Balachandran:1983fg},
who showed that the non-Abelian Coulomb magnetic field creates a
topological obstruction that prevents one from globally defining a
basis for the unbroken color subgroup.  This inability to define
``global color'' is the fundamental reason for the nonexistence of the
SU(5) chromodyons\footnote{The SU(5) monopoles do have dyonic
counterparts with an electric charge in the U(1) subgroup defined by
the magnetic charge, which lies partly in the unbroken
SU(3)~\cite{Nelson:1983em}.  However, because the electric charge is
restricted to this subgroup, implying that one cannot generate full
color multiplets of states, and because the color electric charge is
strictly proportional to the Abelian electric charge, these are
chromodyons only in a limited sense.}.

These barriers to the existence of a chromodyon would both be absent
if the total magnetic charge were purely Abelian.  This can
certainly be achieved by assembling a collection of magnetic monopoles
such that the non-Abelian components of their charges sum to zero; a
two-monopole example of this was studied by Coleman and
Nelson~\cite{Nelson:1983fn}.  However, what we need to produce a
chromodyon is a single monopole with purely Abelian magnetic charge.
While there are no such monopoles in the SU(5) theory, they do exist in a
theory with SO(5) broken to U(1)$\times$SU(2).  These were first
discovered~\cite{Weinberg:1982jh} in the BPS
limit, where they appear as
spherically symmetric classical solutions that are characterized by a
``cloud size'' $b$ that can take on any positive value.  These can be
interpreted as being composed of a massive monopole, carrying both
Abelian and non-Abelian magnetic charge, and a massless monopole, with
only non-Abelian magnetic charge.  At the semiclassical level, the
latter is manifested as a cloud of non-Abelian field, of radius $b$,
that surrounds the core of the massive monopole and completely shields
the non-Abelian part of its magnetic charge.  In the BPS limit the
energy is independent of $b$.  However, if the Lagrangian includes a
nonvanishing potential, the cloud size is no longer arbitrary, but
rather is fixed.  This then gives a magnetic monopole whose long-range
field lies only in the U(1) sector, but whose core transforms under
the unbroken SU(2) and thus gives rise to the gauge zero modes from
which we might hope to construct a chromodyon.  It is this system that
we will study.

We start, in Sec.~\ref{so5section}, by constructing the static SO(5)
monopole.  An analytic solution exists for the BPS case.  However, as
we explain later on, the possibility of varying the cloud size makes
these unsuitable for our purposes.  Instead, we must take for our
starting point a non-BPS monopole, for which the field equations must
be solved numerically.  Then, in Sec.~\ref{sec:simInitialData}, we
construct a chromodyonic configuration from this monopole by applying
an SU(2) gauge rotation and solving for the $A_0$ field that is
required by Gauss's law.  Although this is not an exact static
solution of the field equations, one might expect it to be close to
the desired chromodyon.  We test this by numerically simulating the
time evolution with this as the initial configuration.  We describe
the details of this simulation in Sec.~\ref{evolutionsection}.  The
results are described in Sec.~\ref{slowdownsection}.  We find that,
rather than evolving toward a stable chromodyon, the chromodyonic
configuration continually radiates non-Abelian charge.  Although we
are not able to continue the simulation long enough to verify that
this charge is completely radiated away, every evidence indicates that
this will be the case.  Our conclusions are summarized in
Sec.~\ref{conclusion}.  There are two Appendices containing some
technical details.

\section{SO(5) monopoles}
\label{so5section}

We are interested in theories with Lagrangian densities of the form 
\begin{equation}
    {\cal L} = -{1 \over 4} \Tr F^{\mu\nu}F_{\mu\nu} 
  -{1 \over 2}\Tr D^\mu \Phi \,D_\mu \Phi - V(\Phi)  \, .
  \label{eqn:lagden}
\end{equation}
Here the gauge field $A_\mu$ and the adjoint representation Higgs
field $\Phi$ are both written as imaginary antisymmetric $5 \times 5$
matrices.

To describe the components of these and other adjoint representation
fields, we will adopt the following conventions.  In the defining
representation, the generators of SO(5) are the ten 5$\times$5 matrices
\begin{equation}
(J^{mn})_{ij}= - i(\delta_{im}\delta_{jn}-\delta_{in}\delta_{jm})
           \, ,  \qquad   1 \le m < n \le 5 \, .
\end{equation}
From these we can define six matrices 
\begin{equation}
   h_a = {1\over 2}\left( {1\over 2}\epsilon_{abc} J^{bc} + J^{a4} 
      \right) \, , \qquad 
   k_a = {1\over 2}\left( {1\over 2}\epsilon_{abc} J^{bc} - J^{a4} 
     \right)
\label{hANDkdef}
\end{equation}
that generate SO(4)=SU(2)$\times$SU(2).  
We can then decompose any adjoint representation field $P$ in 
terms of two triplets, $P_{(1)}^a$ and $P_{(2)}^a$, and  
$P_{(3)}^\mu$ ($\mu=1,2,3,4$) via
\begin{equation}
    P = {\bf P}_{(1)}{\bf \cdot  h} + {\bf P}_{(2)} {\bf\cdot  k}
         + P_{(3)}^\mu J^{\mu 5} \, .
\label{eqn:ericknotation}
\end{equation}
We will refer to $P_{(1)}$, $P_{(2)}$, and $P_{(3)}$ as the first-,
second-, and third-sector components, respectively.

SO(5) can be broken to SU(2)$\times$U(1) in two inequivalent ways.  In
the first, corresponding to the decomposition
${\rm SO(5)}\supset{\rm SO(3)}\times{\rm SO(2)}$, the SU(2) is the subgroup, with
the generators $J^{ab} = \epsilon_{abc}(h_c +k_c)$, that rotates the
first three components of a five-vector among themselves.  We will be
concerned with the second possibility, in which the unbroken SU(2) is
one of the factors of the ${\rm SO}(4) = {\rm SU(2)}\times{\rm SU(2)}$
subgroup that mixes the first four components of a five-vector among
themselves; we will choose it to be the subgroup generated by the
$k_a$.  

We will be seeking spherically symmetric monopole solutions.  If we
also require that the fields have positive
parity, the most general spherically symmetric ansatz can be written
as\footnote{In Ref.~\cite{Weinberg:1982jh}, the third-sector fields
were written in a somewhat different manner than here.  The
normalizations in the present ansatz have been chosen so that the
coefficient functions are, nevertheless, the same as those that appear
in that paper.}
\begin{eqnarray}
A^a_{i(1)} = \epsilon_{aim} {\hat r}_m A(r) \, , && \quad\quad
\Phi^a_{(1)} = {\hat r}_a H(r) \, ,\hspace*{0.8in}  \cr
A^a_{i(2)} = \epsilon_{aim} {\hat r}_m G(r) \, ,&& \quad\quad \Phi^a_{(2)}
= {\hat r}_a K(r) \, , \cr
A_{i(3)}^\mu = \sqrt{2} \left[\delta^{i\mu}\, F(r) 
    + \delta^{\mu a}{\hat r}_i {\hat r}_a S(r)\right] \, ,
     && \quad\quad \Phi^\mu_{(3)} = -\sqrt{2} \,\delta^{4\mu}J(r) \, ,
\label{ansatz}
\end{eqnarray}
where Latin indices run from 1 to 3 and $\mu$ runs from 1 to 4.

Actually, there is some redundancy in this ansatz.  A gauge 
transformation of the form 
\begin{equation}
    \Lambda_S = e^{i \psi(r) \hat r^a J^{a5}}
\label{LambdaSdef}
\end{equation}
preserves the ansatz, but with new coefficient functions, which we
indicate with a tilde, given by 
\begin{eqnarray}
  \tilde H + \tilde K &=& H+K \, , \cr  
  \tilde H - \tilde K &=& (H-K)\cos \psi + 2\sqrt{2} J\sin \psi \, ,\cr
  2\sqrt{2} \tilde J &=&  -(H-K) \sin \psi + 2\sqrt{2} J\cos \psi \, , \cr
   \tilde A - \tilde G &=& A-G  \, ,\cr
   \tilde A + \tilde G + {2\over er} &=& \left(A+G + {2\over er}\right)
          \cos\psi    + 2\sqrt{2} F\sin \psi  \, ,\cr
    2\sqrt{2} \tilde F &=& -  \left(A+G + {2\over er}\right) \sin\psi
              + 2\sqrt{2} F\cos \psi  \, , \cr
     \tilde S &=& S + F (1 -\cos \psi)  + {1\over 2\sqrt{2}} 
           \left(A+G + {2\over er}\right)\sin\psi 
           - {1 \over \sqrt{2}\, e} {d\psi \over dr} \, .
\label{SgaugeTrans}
\end{eqnarray} 
From the last of these equations, we see that $S(r)$ can always be gauged away
with a suitable choice of $\psi(r)$.  We will henceforth assume that this 
has been done, so that $S(r)$ vanishes identically.

Requiring that the fields be nonsingular at the origin gives the boundary 
conditions
\begin{equation}
    A(0)=G(0)= H(0)=K(0) =0 \, .
\end{equation}
The functions $F$ and $J$ can be nonzero at the origin.  However, examination
of the field equations, which we will display below, shows that nonsingular solutions
must have 
\begin{equation}
     F'(0) = J'(0) = 0 \, ,
\end{equation}
where a prime denotes differentiation with respect to $r$.

To obtain the symmetry breaking that we want, the asymptotic value of the
Higgs field must lie in the subgroup generated by the $h_a$, giving the boundary 
conditions
\begin{equation}
     H(\infty) = v \, , \qquad K(\infty)=J(\infty)=0  \, .
\label{scalarATinfinity}
\end{equation}
With this choice, the third-sector gauge fields are massive, so
$F(r)$ falls exponentially fast at large distance.  The behavior of
the other, massless, gauge fields depends on the magnetic charge.  If
the latter is a purely Abelian unit charge, then at large distance  
\begin{equation}
    A(r) \sim -1/er \, , \qquad G(r) \lesssim {\rm const}/r^2 \, , 
         \qquad F(r) \sim e^{-evr/2}  \, .
\label{AatInfinity}
\end{equation}

In Ref.~\cite{Weinberg:1982jh} it was shown that in the BPS limit of vanishing scalar 
potential there is a solution given by 
\begin{eqnarray}
&& A(r) = { v \over \sinh  ev r } - \frac{1}{er} \,
   , \cr \cr
&&  H(r) = v \, \coth  ev r - \frac{1}{er} \, , \cr \cr
&& G(r) = K(r) = \left({v \over \sinh  ev r}-{1 \over  er}\right) \, 
         L(r; b) \, ,\cr \cr
&& F(r) = -J(r) = {v \over \sqrt{8} \, \cosh  (ev r/2)} \, \sqrt{L(r; b)} \, ,
\cr \cr
&& S(r) = 0  \, ,
\label{SOfiveSoln}
\end{eqnarray}
where $b$ is any positive real number and
\begin{equation}
L(r; b) = \frac{b}{b+r \, \coth (ev r/2)} \, .
\end{equation}

This solution can be interpreted as being composed of two distinct
fundamental monopoles.   One is a massive monopole, with core radius
$\sim 1/ev$, whose magnetic charge has both non-Abelian and Abelian
components.   The other is a massless monopole that is manifested at
the semiclassical level as a cloud of radius $b$ whose magnetic charge
cancels the non-Abelian part of the massive monopole's charge.  This 
can be seen by computing the large distance behavior of the magnetic 
field.  For $1/ev \ll r \ll b$, both $A(r)$ and $G(r)$ fall as $1/r$,
and so
\begin{equation} 
      B_{i(1)}^a = {{\hat r_a}{\hat r_i} \over er^2}
      + O(1/r^3)\, ,
    \label{B1asymp}
\end{equation}
\begin{equation}
B_{i(2)}^a = {{\hat r_a}{\hat r_i} \over er^2} 
      + O(1/r^3)\, .
   \label{B2asymp}
\end{equation}
(The third-sector fields fall exponentially outside the massive
monopole core and play no role here.)  Outside the massless cloud, $r
\gg b$, $G(r) \sim -b/er^2$.  As a result, $B_{i(2)}^a$ falls faster
than $1/r^2$, while $B_{i(1)}^a$ is unchanged.  Hence, the long-range
magnetic field, and thus the total magnetic charge, have only 
first-sector components and are purely Abelian.  

One might think that the knowledge of the explicit analytic form of
this monopole solution would make it an ideal case for constructing
chromodyons.  As we will see in the next section, this turns out not
to be so.  The difficulty arises from the fact that the
energy of the BPS monopole is independent of the cloud size $b$, so
that a small perturbation can cause the cloud to expand without bound.
To avoid this problem, we will add a potential term that effectively
fixes the cloud size.  With this term included, the BPS limit no 
longer applies, so we will have to solve the full set of second-order
field equations.  Because this cannot be done analytically, we will
resort to numerical solution.

Thus, let us add a potential of the form 
\begin{equation}
V(\Phi) = -{\mu^2 \over 2} \mbox{Tr}\, \Phi^2 + a(\mbox{Tr}
\,\Phi^2)^2 + b\mbox{Tr}\,\Phi^4 
\label{nonbpsv}
\end{equation}
where $\mu$, $a$, and $b$ are constants.  In order to obtain the desired
symmetry breaking, $b$ must be positive, while the requirement that 
the potential be bounded from below gives the condition $4a+b>0$.  At the 
minimum of the potential, 
\begin{equation}
\mbox{Tr}\, \Phi^2 \equiv v^2 = { \mu^2 \over 4 a + b } \, .
\label{eqn:VminCond}
\end{equation}
There is an SU(2) singlet Higgs scalar with mass
\begin{equation}
   m_s = \sqrt{2} \, \mu
\end{equation}
and an SU(2) triplet with mass
\begin{equation}
   m_t = \sqrt{2(1-c)}\, \mu   \, ,
\end{equation}
where
\begin{equation}
     c \equiv {4a \over 4a+b} \, .
\end{equation}
The positivity of $b$ implies that $-\infty < c < 1$, with the two limits
corresponding to $m_s \ll m_t$ and $m_t \ll m_s$, respectively.  

Substitution of our spherically symmetric ansatz, Eq.~(\ref{ansatz}), into
the Euler-Lagrange equations,
\begin{eqnarray}
\nonumber D_j F^{ji} &=& ie[\Phi, D^i \Phi], \\[5pt]
D_j D^j \Phi &=& {\partial V \over \partial \Phi } \, ,
\label{eqn:eomg}
\end{eqnarray}
yields seven ordinary differential equations (ODEs), corresponding to
the seven coefficient functions in our ansatz.  Note that even though
we can use the gauge freedom described by Eqs.~(\ref{LambdaSdef}) and
(\ref{SgaugeTrans}) to make $S(r)$ identically zero, there is still a
corresponding ODE.  However, while the other six ODEs are second
order, this last is a constraint equation relating the
coefficient functions and their first derivatives.  If we set $S=0$,
this equation, which we will refer to as the $S$ constraint, takes the
form
\begin{eqnarray}
   0 &=& \left(A+G+{2\over er} \right) F' - \left(A'+G' -{2\over er^2}\right) F
     +{1\over 2} (H-K)J' - {1\over 2}(H'-K')J   \cr 
     &&
     -e\left[{1\over 2} \left(A+G+{2\over er} \right)^2 
   +{1\over 4}(H-K)^2 + 2J^2 +4F^2 \right]F
\end{eqnarray}
while the other six ODEs become \newpage
\begin{eqnarray}
  A''&=&-\frac{2}{r}A'+\frac{2}{r^2}A+6eFF'+\frac{3e}{r}A^2+\frac{e}{r}H^2
      \cr &&+e^2 \left(A^3 +AH^2 +AJ^2 +5AF^2 -GF^2 -GJ^2 +3HFJ +KFJ \right)
     \,, \cr \cr
   G''&=&-\frac{2}{r}G'+\frac{2}{r^2}G-6eFF'+\frac{3e}{r}G^2+\frac{e}{r}K^2
     \cr && + e^2 \left(G^3
     +GK^2 +GJ^2+5GF^2-AF^2-AJ^2-3KFJ-HFJ \right) 
   \,, \cr\cr
   F''&=&-\frac{1}{r}F'+\frac{e}{2}(A+G)F'+ e(A' +G')F
    +\frac{2e}{r}(A+G)F  +\frac{e}{2r}(H-K)J
     \cr && +e^2 \left(4F^2+ A^2+ G^2 
 +\frac{1}{4}H^2+\frac{1}{4}K^2+2J^2+\frac{1}{2}HK-AG\right)F 
   \cr && +  \frac{e^2}{4}(3AH-3GK-GH +AK)J        
   \,,  \cr\cr
     H''&=&-\frac{2}{r}H'+\frac{2}{r^2}H-4eFJ'-2eF'J
   +\frac{4e}{r}AH 
       \cr 
   &&+ e^2 \left(3F^2 H + F^2 K+2A^2 H+6AFJ-2FGJ \right) \cr
    && + {\mu^2\over v^2} \left[H(H^2 + 3K^2 -v^2) +4J^2(H-K) +c(4J^2K-2HK^2) \right]
      \,,  \cr\cr
  K''&=&-\frac{2}{r}K'+\frac{2}{r^2}K+4eFJ'+2eF'J
     +\frac{4e}{r}GK+
    \cr &&+ e^2\left(2G^2 K+3F^2 K+F^2 H+2AFJ-6GFJ \right) \cr
    &&+{\mu^2\over v^2} \left[ K(K^2+3H^2 -v^2) + 4J^2(K-H) -c(2H^2K - 4J^2H) \right]
     \,,  \cr\cr
   J''&=&-\frac{2}{r}J'+\frac{e}{2}HF'-\frac{e}{2}KF'+eH' F-eK'F
    + \frac{2e}{r}HF-\frac{2e}{r}KF 
   \cr && +{e^2\over 2}\left(A^2 + G^2 +12F^2 -2AG\right)J 
    + {e^2\over 2}\left(3AH -3GK +AK -GH \right)F  \cr
     && +{\mu^2\over v^2} \left[ 8J^2 +(H-K)^2 -v^2 +c(2HK -4J^2) \right]J
    \, .
\label{eqn:seom}
\end{eqnarray}

These equations are not all independent.  For example, the $F''$ equation can
be derived from the $S$ constraint and the other five ODEs.  The
converse is not quite true, because there are solutions of the six
second-order equations that do not satisfy the $S$ constraint.
However, if the $S$ constraint holds at one value of $r$, the
remaining ODEs imply that it holds for all $r$.  In particular, the
$S$ constraint is satisfied for all $r$ if the fields at spatial infinity
obey Eqs.~(\ref{scalarATinfinity}) and (\ref{AatInfinity}).

In the BPS case, the exponential approach of the coefficient functions
to their asymptotic behavior is governed by a single mass scale, $ev$.
With the potential added, three different mass scales --- $ev$, $m_s$,
and $m_t$ --- come into play.  To simplify our numerical simulations
of the time evolution and to avoid the well-known stiffness problem,
we want these characteristic lengths to be close to each other.  To
that end we set $\mu = ev$ and $c=0.5$, so that $ev = m_t =
m_s/\sqrt{2}$.

\begin{figure}
\begin{center}
\leavevmode
\epsfysize=5in
\epsffile{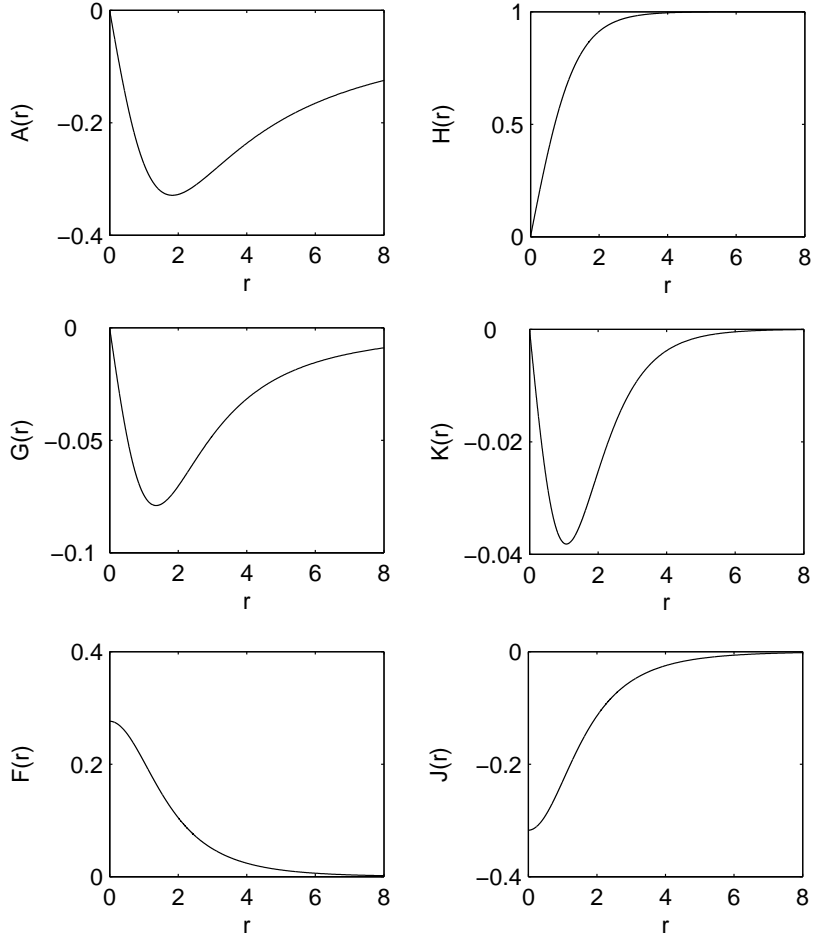}
\end{center}
\caption{Monopole solution in the non-BPS SO(5) gauge theory.  The 
radial distance $r$ is given in units of $1/ev$, and the coefficient 
functions in units of $v$.}
\label{fig:nonbpssol}
\end{figure}

We choose to numerically solve the six ODEs in
Eq.~(\ref{eqn:seom}) and then check the solution against the $S$
constraint. We use a MATLAB package -- SBVP 1.0~\cite{sbvp} -- to
solve this boundary value problem. By using the collocation method,
the SBVP numerical package can handle the singular terms in
Eq.~(\ref{eqn:seom}) with high accuracy near the origin, where
$r\rightarrow 0$. Our numerical solution is shown in
Fig.~\ref{fig:nonbpssol}. The $S$ constraint has been checked to be
automatically satisfied within the numerical error.

It should be noted that this solution is not unique.  Setting
$G=F=J=K=0$ reduces the field equations to two coupled equations for
$A$ and $H$ that are identical to those of the SU(2) theory.   This then
yields a solution that is simply an embedding of the SU(2) unit monopole
via the subgroup generated by the $h_a$.  For our choice of parameters, 
the mass of this pure SU(2) solution is $1.287 M_{\rm BPS}$,
where $M_{\rm BPS} = 4\pi v/e$ is the mass of the unit BPS monopole. 
By contrast, the non-embedding solution shown 
in Fig.~\ref{fig:nonbpssol} has a mass of $1.253 M_{\rm BPS}$.
Note that the mass difference is numerically significant, well above
the numerical errors.

\section{Constructing a chromodyonic configuration}
\label{sec:simInitialData}

In a U(1) dyonic soliton the electric charge results from a
time-dependent phase of a complex field.  In a similar fashion, a
magnetically charged configuration with a time-dependent orientation
with respect to a non-Abelian group carries non-Abelian electric
charge and is a chromodyon.  In this section we will show how such
chromodyons can be constructed from time-independent solutions such as
those obtained in the previous section.  We start with a static
solution $\{(A_i)_{\rm static}({\bf r})$, $(\Phi)_{\rm static}({\bf
r})\}$.  As our first step, we excite one of the zero modes of the
static solution and uniformly rotate its SU(2) orientation to obtain
\begin{eqnarray}
\nonumber (A_i)_{\rm I} &=&  R(t) \, (A_i)_{\rm static} R^{-1}(t) \, ,\\
(\Phi)_{\rm I} &=& R(t) (\Phi)_{\rm static} R^{-1}(t) \, ,
\label{firstRotation}
\end{eqnarray}
where
\begin{equation}
R(t) = e^{i\,k_3\omega_0 t} 
\label{rotationGauge}
\end{equation}
and the generator $k_3$ is defined in Eq.~(\ref{hANDkdef}).  It is critical to
realize that this is not a gauge transformation, because 
the latter would have required that we also add a spatially constant $A_0
= -(\omega_0/e) k_3$.  However, we do need a nonzero $A_0$ in order to satisfy
the Gauss's law constraint
\begin{equation}
   D_j F^{j0} = ie[\Phi, D^0\Phi]  \, .
\label{GaussLaw}
\end{equation}
Solving this equation, given $(A_i)_{\rm I}$ and $(\Phi)_{\rm I}$ and
the boundary condition
\begin{equation}
    A_0(\infty) = 0 \, ,
\end{equation}
yields a solution that we denote by $(A_0)_{\rm II}$.  This gives us a
time-dependent configuration $\{(A_0)_{\rm II},(A_i)_{\rm I},
(\Phi)_{\rm I}\}$.

It is often more convenient to work instead with the stationary
configuration obtained by applying a gauge transformation with gauge
function $\Lambda = R^{-1}$.  This gives us $\{(A_0)_{\rm III},
(A_i)_{\rm static},(\Phi)_{\rm static}\}$ where
\begin{equation}
   (A_0)_{\rm III} =  {\omega_0\over e}\,k_3 + R^{-1} (A_0)_{\rm II} R \, . 
\end{equation}
It is not hard to see that we could have obtained this final
configuration by directly solving Gauss's law with $(A_i)_{\rm
static}$ and $(\Phi)_{\rm static}$ given and $A_0$ obeying
a different boundary condition,
\begin{equation}
  A_0(\infty)  =  {\omega_0\over e}\,k_3 \,.
   \label{eqn:so5chromodyona0bc}
\end{equation}
We denote the solution of this equation as $(A_0)_{\rm static}$.  

Even when working in this static gauge, it is convenient to describe the 
extra energy associated with this configuration in rotational terms.   The 
relevant ``moment of inertia'' is given by the spatial integral of the 
sum of the squares of the field components that undergo the phase rotation.
Since the characteristic length scale is $\sim (ev)^{-1}$, while the natural
scale of the fields is $v$, we have
\begin{equation}
    E_{\rm ch} \sim I \omega_0^2 \sim \left({1 \over ev}\right)^3 v^2 
     \omega_0^2 \sim {v \over e} \left({\omega_0 \over ev}\right)^2 \, .
\label{energyOFrot}
\end{equation}
Similarly, the SU(2) electric charge has a magnitude
\begin{equation} 
    q_E \sim e I \omega_0 \sim {1 \over e} \left({\omega_0 \over ev}\right)
\end{equation}
with the extra factor of $e$ arising because the electric charge is $e$ times
the momentum conjugate to the phase rotation.

The configuration $\{(A_0)_{\rm static},(A_i)_{\rm static},(\Phi)_{\rm
static}\}$ has both a U(1) magnetic charge and an SU(2) electric
charge, and so is a chromodyon.  The question that we need to address
is whether it is a solution of the field equations.  It is easy to see
that it cannot be, because the rotation in group space induces a
deformation of the field profiles, just as the spatial rotation of a
solid object induces a deformation of its shape.  However, this
deformation is\footnote{This can be seen easily from the field
equations
\begin{eqnarray}
\nonumber D_0 F^{0i} + D_j F^{ji} = ie[\Phi, D^i \Phi] \, , \\
D_0 D_0 \Phi - D_j D_j \Phi = 0\, ,
\label{eqn:staticEOM}
\end{eqnarray}
where the non-static terms, $D_0 F^{0i}$ and $D_0 D_0 \Phi$, are of
order $O(\omega_0^2)$. If these terms are omitted, the equations
reduce to the static equations satisfied by $(A_i)_{\rm static}$ and
$(\Phi)_{\rm static}$.}  of order $\omega_0^2$, and so should require
only a small modification of the configuration if $\omega_0$ is
sufficiently small.  If this is the only correction needed, then the
theory does indeed have a stable static chromodyon solution.  On the
other hand, it may be that there is no such static solution.  We will test
for this possibility by taking $\{(A_0)_{\rm
static},(A_i)_{\rm static},(\Phi)_{\rm static}\}$ as an initial
condition and then letting the fields evolve in time according to the
field equations.  If the deformation resulting from the rotation in
SU(2) space is the only impediment to its being a static solution, the fields
should oscillate, with an initial amplitude proportional to $\omega_0^2$, 
and eventually settle down in 
the true static solution. 

In order to do this, we need to determine $(A_0)_{\rm static}$.  Thus,
our immediate task is to solve Eq.~(\ref{GaussLaw}) subject to the boundary
condition that $A_0(\infty) =  (\omega_0/e)k_3$.  The fields 
$(A_i)_{\rm static}$ and $(\Phi)_{\rm static}$ are both spherically 
symmetric.  However, the boundary condition on $A_0$ breaks this symmetry,
so we can only assume that $A_0$ has an axial symmetry.  The most general 
ansatz for $A_0$ is then 
  \begin{eqnarray}
    && A_{0(1)}^a = {\hat r}^a u(\rho,z) \, ,
      \qquad\quad A_{0(1)}^3 = w(\rho,z) \, , \cr 
   &&   A_{0(2)}^a   = {\hat r}^a b(\rho,z) \, ,
       \,   \qquad\quad A_{0(2)}^3 = Q(\rho,z) \, , \cr
    &&  A_{0(3)}^a = \sqrt{2}\,\epsilon_{ab} {\hat r}^b q(\rho,z) \, , 
      \quad\quad A_{0(3)}^3 = 0 \, ,
     \cr &&  A_{0(3)}^{4} = -\sqrt{2}\,t(\rho,z)  \, ,
  \label{eqn:a0axialansatz}
  \end{eqnarray}
where $a$ and $b$ are either 1 or 2 and $\rho=\sqrt{x^2+y^2}$.  The
boundary conditions at spatial infinity require that $Q$ approach
$\omega_0/e$ and that the other coefficient functions tend to zero.

Substituting this ansatz into Gauss's law, Eq.~(\ref{GaussLaw}),
yields the set of second-order partial differential equations that are
displayed in Appendix~\ref{apx:a0gausseqn}.  These rather complicated
equations experience a remarkable simplification if $(A_i)_{\rm
static}$ and $(\Phi)_{\rm static}$ are taken to be the BPS solution of
Eq.~(\ref{SOfiveSoln}).  In this case, they can be satisfied by
setting all the coefficient functions except $Q$ to zero, taking
$Q(\rho,z)=Q(r)$, and requiring that
\begin{equation}
  \frac{d Q(r)}{d r} + e G(r) Q(r) = 0 \, ,
  \label{eqn:bpsgausseqn}
\end{equation}
where $G(r)$ is the second-sector gauge field function given in
Eq.~(\ref{SOfiveSoln}).  The solution to Eq.~(\ref{eqn:bpsgausseqn})
is\footnote{This result was also obtained, by a different method,
in Ref.~\cite{Lee:1996vz}.}
\begin{equation}
  Q(r) = {\omega_0 \, b \over e L(r,b)} \, 
        {\partial L(r,b) \over \partial b} = \left({\omega_0 \over e}\right)
   \frac{r \, \mbox{coth}(evr/2)}{b+r \, \mbox{coth}(evr/2)} \, .
\end{equation}

Unfortunately, this simple solution turns out not to be useful.  To
see this, recall that for small velocities the dominant
time dependence of a soliton arises entirely through
excitation of 
its zero modes, whose dynamics is governed by the moduli space
Lagrangian.  For the generic case, with a nonzero scalar field
potential, there are seven zero modes about $\{(A_i)_{\rm
static},(\Phi)_{\rm static}\}$.  Four of these --- three translation
modes and one U(1) phase mode --- are irrelevant for our purposes.
The remaining three are SU(2) orientation modes, one of which has been
excited by the transformation in Eq.~(\ref{firstRotation}).  Within
the moduli space approximation (MSA) there would be uniform motion in
the corresponding collective coordinate.  In the gauge where
$A_0(\infty)=0$, the soliton would rotate uniformly in SU(2) space, as
described by Eq.~(\ref{firstRotation}).  

In the BPS case there is an additional zero mode, corresponding to the
freedom to vary the cloud radius $b$.  The moduli space Lagrangian
governing the eight zero modes is given by
\begin{equation}
  L_{\rm MS} = {1\over 2} M \,\dot{\bf X}^2 
  + {1 \over 2 M } \dot \chi^2 
  + {1 \over 2} \left\{ 
   {\dot b^2 \over b} + b\left[ 
  \dot \alpha^2 + \sin^2\alpha\:\:\dot\beta^2 
  + (\dot\gamma + \cos\alpha\:\,\dot\beta)^2 \right] \right\} \, ,
\end{equation}
where $M$ is the BPS monopole mass, $\bf X$ is the location of the
center of the system, $\chi$ is the U(1) phase, and $\alpha$, $\beta$,
and $\gamma$ are SU(2) Euler angles.  We are interested in the part
within the curly braces, $L_{\rm MS}^{\rm rel}$, that describes the
zero modes corresponding to the non-Abelian cloud size and the SU(2)
orientation of the non-Abelian cloud.  By transforming to coordinates
\begin{eqnarray}
  x_1 &=& 2\sqrt{b} \, \mbox{sin}{\alpha \over 2}\, 
     \mbox{cos}{\, \beta - \gamma \over 2} \, , \cr
    x_2 &=& 2\sqrt{b} \, \mbox{cos}{\alpha \over 2}\,
   \mbox{cos}{\, \beta + \gamma \over 2} \, ,\cr
    x_3 &=& \,2\sqrt{b} \, \mbox{sin}{\alpha \over 2}\,
     \mbox{sin}{\, \beta - \gamma \over 2} \, ,\cr
     x_4 &=& \,2\sqrt{b} \, \mbox{cos}{\alpha \over 2}\,
          \mbox{sin}{\, \beta + \gamma \over 2} \, ,
\end{eqnarray}
we see that this is actually the Lagrangian for a free particle in
four-dimensional Euclidean space,
\begin{equation}
    L_{\rm MS}^{\rm rel} =  {1\over 2} \dot x_1^2 
    + {1\over 2}\dot x_2^2 + {1\over 2}\dot x_3 ^2 
     + {1\over 2}\dot x_4^2  \, .
\end{equation}

The solutions of this Lagrangian are uniform straight-line motion.
Without loss of generality, we focus on solutions in the $x_1$-$x_2$
plane.  The SU(2) rotating configurations we are studying then
correspond to taking initial values $\alpha = \beta = \gamma =0$,
$b=b_0$, $\dot \alpha = \omega_0$, and $\dot \beta = \dot \gamma =
\dot b =0$.   This leads to 
\begin{eqnarray}
   x_1(t) &=& v_0 t \, ,\cr 
   x_2(t) &=& \rho_0 \, ,
\end{eqnarray}
or, equivalently,
\begin{eqnarray}
    b(t) &=& b_0 + b_0 \left( {\omega_0 \, t \over 2} \right) ^2 \, ,\cr
   \alpha(t) &=& 2 \,\, \mbox{tan}^{-1} \, { \omega_0 \, t \over 2 } \, .
\label{MSAforBPS}
\end{eqnarray}
We see that the SU(2) phase does not even go through a full rotation,
so this hardly a good approximation to a uniformly rotating
configuration of fixed color-electric charge.  This is clearly
attributable to the fact that the cloud size can grow without
limit.\footnote{Equation~(\ref{MSAforBPS}) implies that $\dot b$ increases
linearly with time.  In actual fact, the MSA breaks down when when
$\dot b$ approaches the speed of light~\cite{Chen:2001qt}.}

To avoid this difficulty, we turn to the case with a nonzero
potential, where the $b$ mode is no longer a zero mode and the MSA
predicts uniform rotation of the SU(2) orientation.  Because analytic
results are no longer possible, we must resort to numerical solution
of the field equations, both to obtain the static monopole solution,
as described in the previous section, and to find the $A_0$ that
solves the Gauss's law constraint, the topic to which we now turn.

\begin{figure}[t]
\begin{center}
\leavevmode
\epsffile{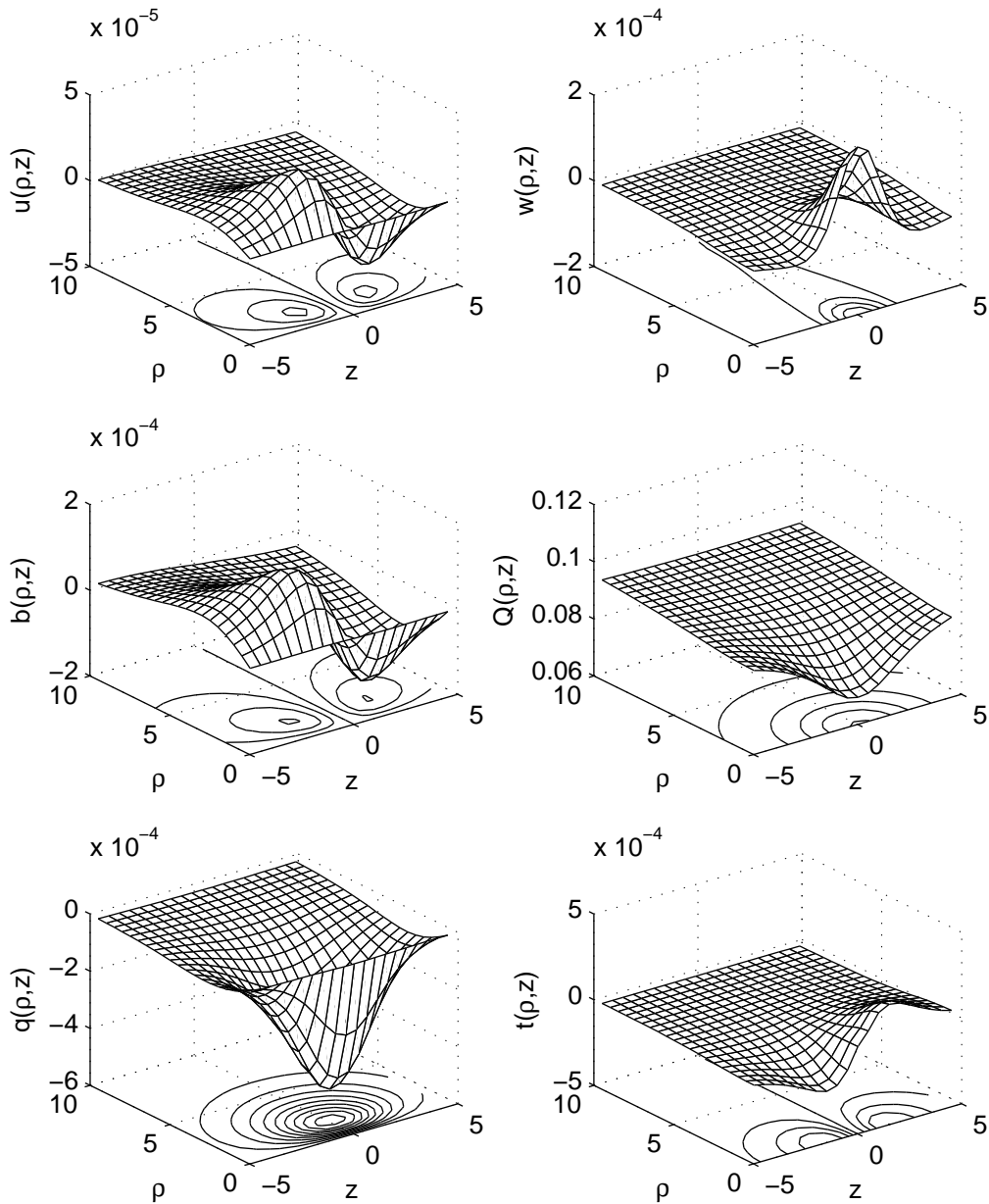}
\end{center}
\caption{Numerical solution for $(A_0)_{\rm static}$ in the non-BPS
SO(5) theory, with $\omega_0 = 0.1\,ev$.  The coordinates $z$ and
$\rho$ are given in units of $1/ev$ and the
component fields $u$, $w$, $b$, $Q$, $q$, and $t$, defined in
Eq.~(\ref{eqn:a0axialansatz}), in units of $v$.}
\label{fig:nonbpsa0ns}
\end{figure}

The ansatz for $A_0$ was given in Eq.~(\ref{eqn:a0axialansatz}), and
the coupled field equations that follow from this ansatz are given in
Appendix~\ref{apx:a0gausseqn}.  The outer boundary conditions are
found by noticing that well outside the core [i.e, when $r$ is much greater than
$(ev)^{-1}$,
$m_{\rm s}^{-1}$, and $m_{\rm t}^{-1}$] the third-sector components are
all exponentially small and we have
\begin{eqnarray}
   \nonumber u(\rho,z) &\rightarrow& {c_1\rho z \over r^4} \, , \\
   \nonumber w(\rho,z) &\rightarrow& {c_1 z^2 \over r^4} \, , \\
   \nonumber b(\rho,z) &\rightarrow& {d_1\rho z \over r^4} \, , \\
    Q(\rho,z) &\rightarrow& \omega_0 +{c_2\over r} + {d_2\over r^2} 
    + {d_1 z^2 \over r^4} \, .
\end{eqnarray}
Here $c_1$ and $c_2$ are free constants, to be determined from the numerical simulation,
while $d_1$ and $d_2$ can be derived in terms of these 
by analysis of the asymptotic expansion.
We use the successive over-relaxation (SOR) method with
red-black ordering as our numerical method. Our results are shown in
Fig.~\ref{fig:nonbpsa0ns}.

In contrast with the BPS case, we see that all of the coefficient functions
are nonzero, although $Q$ remains dominant, and that all of these functions,
including $Q$, have only axial symmetry, with separate dependence
on $\rho$ and $z$.

\section{Evolving the chromodyon}
\label{evolutionsection}

In the previous two sections we obtained a static monopole
solution and then determined the $A_0$ that is required
by Gauss's law when this solution 
rotates uniformly in SU(2) space.  We now take this
configuration as the initial condition and let the system evolve as
dictated by the equations of motion.  We work in the gauge where the
uniform rotation has been gauged away, so that the initial
configuration $\{(A_0)_{\rm static},(A_i)_{\rm static},(\Phi)_{\rm
static}\}$ would be a static solution if the MSA were exact.  In this
gauge, any time dependence arises from corrections to the MSA.

To proceed, we need to specify $\omega_0$.  It cannot be too big
(e.g., so large that the energy arising from the gauge rotation is
comparable to the monopole mass) if the original configuration is to
be even an approximate solution.  A more stringent condition is
suggested by the existence of the embedded pure SU(2) monopole
described at the end of Sec.~\ref{so5section}.  In order to make sure
that our configuration does not evolve toward this other monopole
solution, we want the energy associated with the gauge rotation to be
less than the mass difference between the two types of static monopoles.
On the other hand, taking $\omega_0$ to be too small will impose
increased computational burdens, because we will have to simulate the
evolution for a much longer time in order to see any effect.

We choose $\omega_0=0.04 \, ev$. The gauge rotation energy corresponding to
this is of order $(\omega_0/ev)^2 M_{\rm BPS}\approx 10^{-3}
M_{\rm BPS}$, smaller than the mass difference between the two
monopole solutions.  Because the $A_0$ obtained in the previous
section is linearly proportional to $\omega_0$, the initial data can
be obtained by a simple rescaling of the solution shown in
Fig.~\ref{fig:nonbpsa0ns}.

The Euler-Lagrange equations consist of the evolution equations
\begin{eqnarray}
    && D_0 F^{0i} + D_j F^{ji} = ie[ \Phi, D^i \Phi ]  \, ,
 \label{firstEOM}  \\
&& D_0 D^0 \Phi + D_j D^j \Phi = {\partial V(\Phi) \over \partial \Phi} \, ,
\label{eqn:fulleqn}
\end{eqnarray}
and the Gauss's law constraint
\begin{equation}
D_j F^{j0} = ie[ \Phi, D^0 \Phi ] \, .
\end{equation}
We will use the so-called free evolution scheme to simulate this
constrained system.  In this scheme, we numerically solve the
evolution equations, and use the constraints to monitor the accuracy
of the evolution. It is well known that direct numerical
implementation can have problems with numerical instability.  To 
avoid this, 
we choose to use the technique in
Ref.~\cite{Abrahams:1995nj} to first rewrite the evolution equations in
hyperbolic form. To do this, we take a covariant time derivative of
Eq.~(\ref{firstEOM}) and subtract a covariant spatial derivative of
the Gauss's law constraint, obtaining
\begin{equation}
D^0 D_0 F^{0i} + D^0 D_j F^{ji} - D^i D_j F^{j0} 
   =ie\, D^0  [\Phi, D^i \Phi] - ie\,D^i [\Phi, D^0 \Phi] \, .
\end{equation}
Switching the two covariant derivatives on the left-hand side gives
\begin{equation}
D^0 D_0 F^{0i} + D_j D_0 F^{ij} + D_j D_i F_{j0} + 2ie[ F_{ij}, F_{j0} ] 
   = ie\, D^0 [\Phi, D^i \Phi] - ie\, D^i [\Phi, D^0 \Phi] \, .
\label{eqn:hpeq}
\end{equation}
Next, we take a covariant spatial derivative of the Bianchi identity,
\begin{equation}
D_0 F_{ij} + D_i F_{j0} + D_j F_{0i} = 0 \, ,
\end{equation}
to give
\begin{equation}
D_j D_0 F_{ij} + D_j D_i F_{j0} = D_j D_j F_{i0} \, .
\end{equation}
We substitute this into Eq.~(\ref{eqn:hpeq}) and obtain
\begin{equation}
D^0 D_0 F_{i0} + D^j D_j F_{i0} + 2ie[ F_{ij}, F_{j0} ] 
     = ie\, D^0 [\Phi, D^i \Phi] - ie\,D^i [\Phi, D^0 \Phi] \, .
\label{eqn:hpevolv}
\end{equation}
The full set of equations now  consists of the definition of
$F_{\mu\nu}$ and two wave equations, Eqs.~(\ref{eqn:fulleqn}) and (\ref{eqn:hpevolv}).
Equation~(\ref{firstEOM}) becomes a second constraint.

One advantage of this hyperbolic formulation is that it does not depend
on the gauge condition.   Therefore, by assuming an appropriate 
time-dependent gauge transformation, we can choose $A_0$ to be anything we 
want.  In particular, we will set
\begin{equation}
A_0({\bf r},t>0) = (A_0({\bf r}))_{\rm static}  \, ,
\label{eqn:a0gauge}
\end{equation}
where $(A_0)_{\rm static}$ is the solution shown in
Fig.~\ref{fig:nonbpsa0ns}.  The chromodyon configuration constructed
previously is an approximate stationary solution with this choice of
$A_0$.  If it were exact, Eq.~(\ref{eqn:a0gauge}) would correspond to
choosing a gauge in which the SU(2) rotation was gauged away.  Any
time dependence that we observe will correspond to corrections to this
approximation.

Once $A_0$ has been fixed in this manner, the time-dependent variables
in this formulation are $A_i$, $\Phi$, and $E_i\equiv F_{i0}$.  Their
initial values at $t=0$ are chosen to be $(A_i)_{\rm static}$,
$(\Phi)_{\rm static}$, and $D_i (A_0)_{\rm static}$.  The time
derivatives of $A_i$ and $\Phi$ at $t=0$ are set equal to zero, while
$\partial_0 E_i(t=0)$ is obtained from the constraint
Eq.~(\ref{firstEOM}).  Note that $E_i(t=0)$ is $O(\omega_0)$ and
$\partial_0 E_i(t=0)$ is $O(\omega_0^2)$.

\begin{figure}[tb]
\vspace{9pt}

\centerline{\hbox{ \hspace{-0.25in}
  \epsfxsize=2.0in
   \epsffile{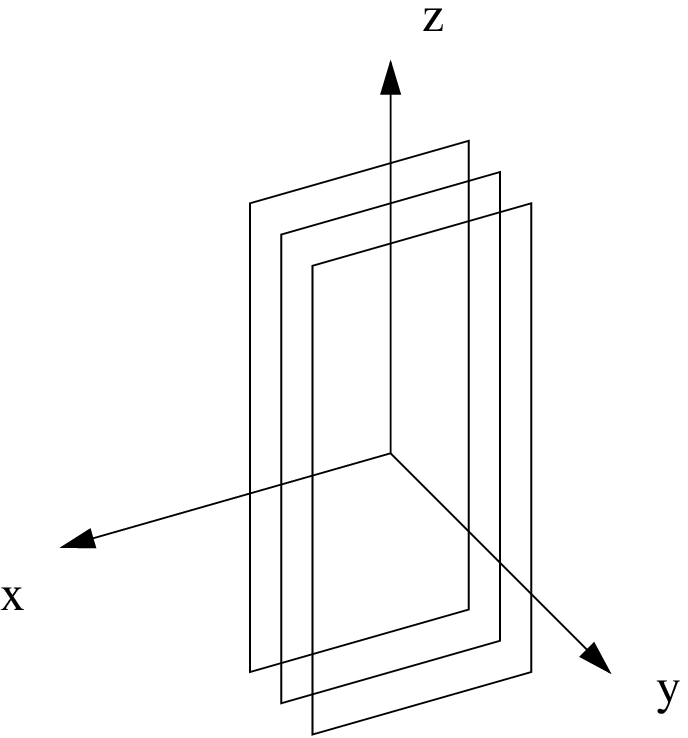}
  \hspace{0.75in}
  \epsfxsize=2.0in
   \epsffile{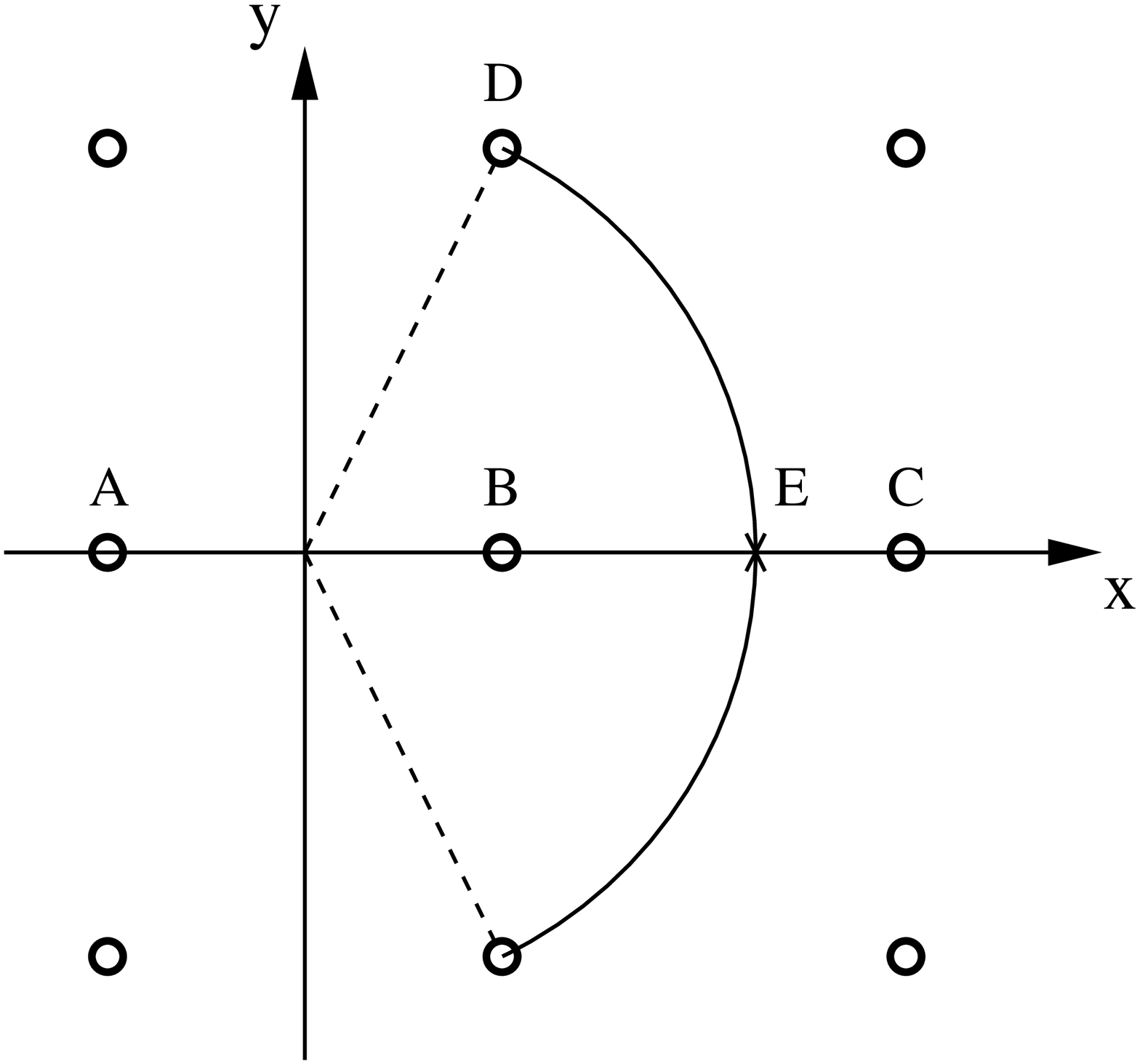}
  }
}

\vspace{9pt}
\hbox{\hspace{1.5in} (a) \hspace{2.5in} (b)}
\vspace{9pt}

\caption{ (a) Using three slabs in the three-dimensional Cartesian
coordinates. (b) Axial symmetry is used to update grid points on
the off-center slabs.}
\label{fig:sandwich_gridstructure}

\end{figure}

The system has axial symmetry at $t=0$ and this symmetry will be
preserved by the evolution. Usually, a two-dimension grid structure in
the $\rho$-$z$ plane is used to discretize an axially symmetric
system. However, using cylindrical coordinates to simulate
time-dependent systems can easily cause numerical
instabilities~\cite{Pogosian:1999zi}. In our implementation, we use
the method proposed in Ref.~\cite{Alcubierre:1999ab}. The idea is to
discretize the system using three slabs in the three-dimension
Cartesian coordinates, as shown in
Fig.~\ref{fig:sandwich_gridstructure}a. Each slab is a two-dimensional
grid structure with step size $h$. The central slab lies in the
$x$-$z$ plane. The other two slabs are obtained by shifting the
central one by $+h$ and $-h$ along the $y$ axis, respectively. At each
grid point, any so(5)-valued element is represented by ten real
numbers. For each time step, the two wave equations are solved first
on the central slab using the finite difference method. Then the axial
symmetry is used to update the grid points on the other two slabs. To
be precise, we plot the grid structure in
Fig.~\ref{fig:sandwich_gridstructure}b. The axial symmetry tells us
that the system will be invariant if we rotate it through the same
angle both in real spatial space and in isospin space. For the $\Phi$
field, this means that
\begin{equation}
\Phi({\bf r}_D,t) = e^{-i h_3 \theta}e^{-i k_3 \theta}\Phi({\bf r}_E,t) 
   e^{ i k_3 \theta}e^{ i h_3 \theta} \, ,
\end{equation}
where $\theta$ is the angle between ${\bf r}_E$ and ${\bf
r}_D$. For the $A_i$ field, it means that
\begin{equation}
A_i({\bf r}_D,t) = e^{-i h_3 \theta}e^{-i k_3 \theta} R_{ij}(\theta) 
  A_j({\bf r}_E,t) e^{ i k_3 \theta}e^{ i h_3 \theta} \, ,
\end{equation}
with $R_{ij}(\theta)$ being the matrix corresponding to a spatial
rotation by angle $\theta$ about the $z$-axis. Interpolation is used
to calculate $\Phi({\bf r}_E,t)$ and $A_i({\bf r}_E,t)$ from the
values at the neighboring grid points, $A$, $B$, and $C$.

In our simulation, the grid with three slabs covers a space of size 60
(in units of $1/ev$).  The open boundary condition is used because of
its simplicity. The radius of the monopole is roughly 1 and we focus
on studying a region with a radius of 10. This implies that we can
only simulate our system up to $t=50$.

Because the system is approximately stationary at
$t=0$, the time-dependent parts are very small. In the hyperbolic
formulation, we have four fields, $A_0$, $A_i$, $E_i\equiv F_{i0}$,
and $\Phi$. Since we have already prescribed $A_0$ to be
time-independent by Eq.~(\ref{eqn:a0gauge}), only $A_i$, $E_i$, and
$\Phi$ have time-dependent parts. In our implementation, we separate
out the time-dependent parts via
\begin{eqnarray}
\nonumber &&A_i({\bf r},t) = {\bar A_i}({\bf r}) + {\tilde A_i}({\bf r},t) \, , \\
\nonumber &&E_i({\bf r},t) = {\bar E_i}({\bf r}) + {\tilde E_i}({\bf r},t) \, , \\
&&\Phi({\bf r},t) = {\bar \Phi}({\bf r}) + {\tilde \Phi}({\bf r},t) \, ,
\label{eqn:evolveperturb}
\end{eqnarray}
where ${\tilde A_i}({\bf r},t=0)={\tilde E_i}({\bf r},t=0)={\tilde
\Phi}({\bf r},t=0)=0$. During the evolution, we check how well these
time-dependent parts satisfy the Gauss's law constraint.
Theoretically, this constraint should be satisfied at $t=0$ because of
the way we construct the initial gauge-rotating system,
and should remain satisfied for all $t$. Numerically,
however, it is only satisfied up to some finite accuracy. To check how
well the Gauss's law constraint is satisfied, we write the right-hand
side of the constraint as a sum of nine terms, and then treat each of
these terms as a vector in a ten-dimensional space.\footnote{This is because
every term in the Gauss's law constraint is in the ten-dimensional
SO(5) adjoint representation.}  We then define the sum of all these
vectors (which theoretically should vanish) to be the defect. This
defect should remain small as long as our numerical scheme is
stable. If the scheme is not stable, the defect will grow
exponentially. We calculate the ratio between the maximum norm of the
defect and the maximum norm among all of its component vectors.  (For
both norms we take the maximum over the entire computational domain.)
We use this ratio to measure the accuracy to which the Gauss's law
constraint is satisfied.  At $t=0$ this ratio is $0.5\times 10^{-2}$,
while during the evolution the ratio for the 
time-dependent parts never exceeds $3\times
10^{-2}$.  More details about this can be found in Appendix
\ref{apx:gausscheck}.

\section{Global gauge rotation slow-down}
\label{slowdownsection}

Our initial configuration satisfies the equations of motion to first
order in $\omega_0$.  If this first-order approximation were exact,
the solution would be time-independent, so the time-dependent parts in
Eq.~(\ref{eqn:evolveperturb}) tell us how the real evolution deviates
from this first-order approximation.  By analyzing the result of our
simulation, we find that we can best fit this deviation in terms of a
global gauge rotation.  Recall that our initial configuration
corresponds to a monopole rotating in SU(2) space about the $k_3$ axis
with an angular velocity $\omega_0$.  We made it static by going to a
gauge with $A_0(\infty) = (\omega_0/e) k_3$, effectively transforming to a
rotating frame.  By fixing $A_0({\bf r},t)$ as in
Eq.~(\ref{eqn:a0gauge}) we stay within that rotating
frame during the evolution.
Any slowing of the gauge rotation would appear in this frame 
as a rotation about the $k_3$ axis, but in the opposite direction.   Thus,
it would correspond to a global gauge rotation generated by 
\begin{equation}
\Lambda_s(\theta) = e^{-i k_3 \theta(t)} 
\label{eqn:smallgt}
\end{equation}
with positive $\theta(t)$.  The effect would be as if
the initial angular velocity $\omega_0$ were replaced by
\begin{equation}
    \omega_{\rm eff} = \omega_0 - {d\theta \over dt}  \, ,
\end{equation}
leading to a configuration with smaller color charge and a smaller 
energy.

The fields $A_i$ and $\Phi$ can be expanded in SO(5) components, as in
Eq.~(\ref{eqn:ericknotation}).  The first-sector components are unchanged
by the rotation generated by $\Lambda_s(\theta)$.   The second-sector
components transform as 
\begin{equation}
\left( \begin{array}{c}
  \hat P_{(2)}^1 \\
  \hat P_{(2)}^2 \\
  \hat P_{(2)}^3 \end{array} \right) =
\left( \begin{array}{ccc}
   \cos \theta & -\sin\theta & 0 \\
   \sin\theta & \cos \theta & 0 \\
  0 & 0 & 1 \end{array} \right)
\left( \begin{array}{c}
  P_{(2)}^1 \\
  P_{(2)}^2 \\
  P_{(2)}^3\end{array} \right) 
\label{eqn:k3gt2}
\end{equation}
while the third-sector components decompose into two doublets
transforming according to
\begin{equation}
\left( \begin{array}{c}
  \hat P_{(3)}^1 \\
  \hat P_{(3)}^2 \\
  \hat P_{(3)}^3 \\
  \hat P_{(3)}^4 \end{array} \right) =
\left( \begin{array}{cccc}
   \cos(\theta/2) & -\sin(\theta/2)  & 0 & 0 \\
  \sin(\theta/2) & \cos(\theta/2) & 0 & 0 \\
  0 & 0 & \cos(\theta/2) & -\sin(\theta/2) \\
  0 & 0 & \sin(\theta/2) & \cos(\theta/2) \end{array} \right )
\left( \begin{array}{c}
  P_{(3)}^1 \\
  P_{(3)}^2\\
  P_{(3)}^3 \\
  P_{(3)}^4 \end{array} \right) \, .
\label{eqn:k3gt3}
\end{equation}

\begin{figure}[th]
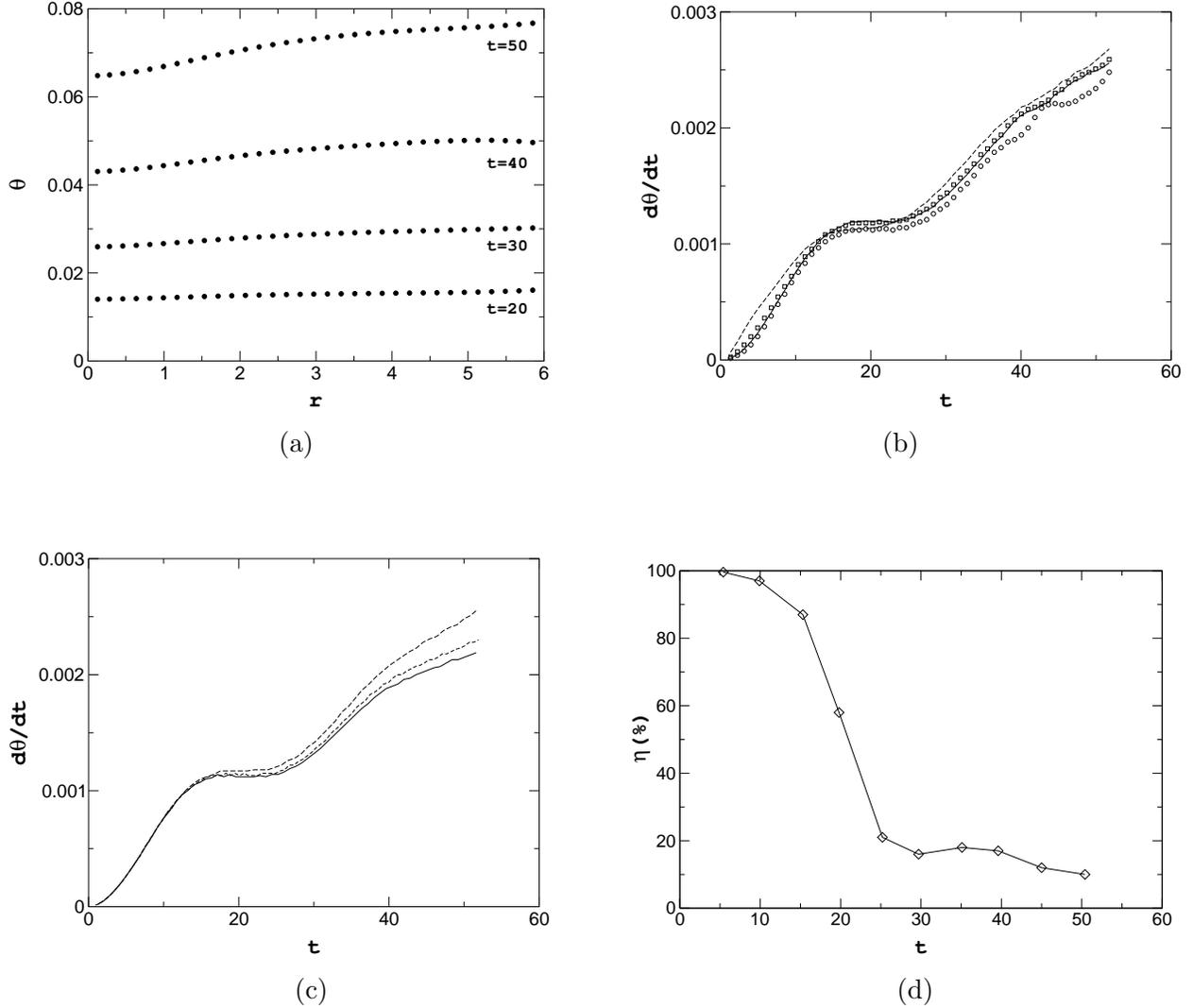

\vspace{9pt}

\centerline{\vbox{\hbox{ \hspace{-0.1in}
  \rotatebox{0}{ \epsfxsize=3.0in \epsffile{figure4a.eps}
}
\hspace{0.3in}
  \rotatebox{0}{ \epsfxsize=3.0in \epsffile{figure4b.eps}
}
  }
\vspace{3pt}
\hbox{\hspace{1.5in} (a) \hspace{3.05in} (b)}
\vspace{.5in}
\hbox{ \hspace{-0.1in}
  \rotatebox{0}{\hspace{0in} \epsfxsize=3.0in \epsffile{figure4c.eps}
}
\hspace{0.3in}
  \epsfxsize=3in {\epsffile{figure4d.eps}
}
}}}
\vspace{3pt}
\hbox{\hspace{1.5in} (c) \hspace{3.05in} (d)}
\vspace{4pt}

\caption{(a) Fitted values of $\theta$ at grid points along the
positive $x$-axis.  (b) Values of $d\theta/dt$ obtained by fitting
with various sets of component fields.  Circles indicate fits using
$\Phi$, squares fits using $A_i$, the solid line fits using
$A_{i(3)}$, and the dashed line fits using $A_{i(2)}$.  (c) Comparison
of results using different grid sizes.  The long dashed curve, the
dashed curve, and the solid line curve are calculated using $\Delta
h=0.18 (ev)^{-1}$, $\Delta h=0.12(ev)^{-1}$, and $\Delta h=0.08
(ev)^{-1}$ grid sizes, respectively.  (d) Plot of $\eta(t)$, defined
by Eq.~(\ref{etaDef}), which indicates the fraction of the time
dependence that cannot be accounted for by the global gauge rotation.
In all four figures distances and times are given in units of $1/ev$
and $\omega_0 =0.04 ev$.  }
\label{fig:theta_fittings}

\end{figure}

Let us define
\begin{eqnarray}
\nonumber \Delta \Phi({\bf r},t;\theta) &=&
\left[\frac{}{}\Lambda_s(\theta) {\bar \Phi}({\bf r})
\Lambda_s^{-1}(\theta) - {\bar \Phi}({\bf r})\frac{}{}\right] -
{\tilde \Phi}({\bf r},t) \, , \\ \Delta A_i({\bf r},t;\theta) &=&
\left[\frac{}{}\Lambda_s(\theta) {\bar A_i}({\bf r})
\Lambda_s^{-1}(\theta) - {\bar A_i}({\bf r})\frac{}{}\right] - {\tilde
A_i}({\bf r},t) \, ,
\label{eqn:apdiff}
\end{eqnarray}
where ${\tilde \Phi}({\bf r},t)$ and ${\tilde A_i}({\bf r},t)$ are
from our numerical simulation.  If the time evolution of our
configuration were completely due to a slowing of the global gauge
rotation, then for any given time $t$ there would be a single 
$\theta(t)$ that would make ${\Delta \Phi}({\bf r},t;\theta)$ and ${\Delta
A_i}({\bf r},t;\theta)$ both vanish for all values of $\bf r$.  To see how 
close we are to this situation, we can extract a value for $\theta$ by 
several different methods and then compare these values.  First, 
we obtain $\theta$ from the second-sector components of $\Phi$
by minimizing
\begin{equation}
     \Tr  \left[ \Delta \Phi_{(2)}({\bf r},t;\theta)\right]^2 
\end{equation}
at various points.  In Fig.~\ref{fig:theta_fittings}a we show the
$\theta$ obtained in this manner for a series of points along the
$x$-axis.  As can be seen, the $\theta$'s thus obtained are only 
weakly position-dependent.
We can also define a spatially averaged $\theta$ by finding the value
that minimizes quantities such as
\begin{equation}
{\cal N}(t, \theta) = \int\!\!\!\int\!\!\!\int_{r\leq R_e}
\left\{\mbox{Tr} \left[\Delta A_i({\bf r},t;\theta) \right]^2
        +\mbox{Tr}\left[\Delta\Phi({\bf r},t;\theta) \right]^2 \right\}dxdydz \, ,
\label{eqn:thetafit}
\end{equation}

\noindent where $R_e$ is the size of the physical region of
interest.\footnote{The numbers we present are obtained using $R_e =
6(ev)^{-1}$, but these results are not very sensitive to the exact
value of $R_e$.}  In Fig.~\ref{fig:theta_fittings}b, we compare
the results for $d\theta/dt$ that are 
obtained by restricting ${\cal N}$ in several different ways:
using only $\Phi$ or only $A_i$, or using just the second-sector or
just the third-sector components of $A_i$.  [The first-sector
components are invariant under $\Lambda_s(\theta)$, and so cannot
affect the fitting of $\theta$.]  We see that all of these methods
give essentially the same results, again consistent with the
interpretation in terms of a spatially uniform global gauge rotation.
In order to indicate the convergence of our simulations, in
Fig.~\ref{fig:theta_fittings}c we show three curves for
$\omega_0=0.04ev$, using successively finer grid structures.

We can also ask how much of the time-dependence can be accounted 
for by this uniform rotation.  To this
end, we divide the residual norm after fitting $\theta$ by the norm of all
the component fields in the time-dependent parts, and define
\begin{equation}
\eta(t) =\frac{{\cal N}(t,\theta(t))}{{\cal N}(t,0)} \, .
\label{etaDef}
\end{equation}
We plot $\eta$ for our simulation in Fig.~\ref{fig:theta_fittings}d.  We
see that, although initially the global gauge rotation accounts for only 
a small part of the time dependence, at large times it is clearly the 
dominant component. 

\begin{figure}
\vspace{9pt}

\centerline{\hbox{ \hspace{-0.1in}
  \rotatebox{0}{ \epsfxsize=3.0in \epsffile{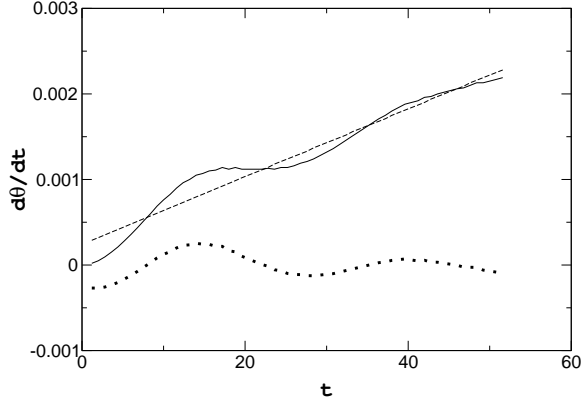}
}
  }
}

\vspace{9pt}
\caption{Decomposition of $d\theta/dt$ (solid line) into linear
and oscillating components. Time is in units of $1/ev$, and $\omega_0=0.04 ev$.}
\label{fig:sim_results}

\end{figure}

We see from the data in Fig.~\ref{fig:theta_fittings} that
$d\theta/dt$ increases with time (although not uniformly),  with a corresponding
decrease in $\omega_{\rm eff}$.  This can be interpreted as the sum of
two effects, as shown in Fig.~\ref{fig:sim_results}.  One is an
overall oscillation pattern that appears to be a transient effect
caused by the relaxation of the system after the initial excitation.
The other effect is a linearly increasing $d\theta/dt$.  By the end of
our simulation, at $t = 52 (ev)^{-1}$, about 5\% of the initial
angular velocity has been lost.  We do not see any indication that
this slowing down process will stop, and expect that $\omega_{\rm eff}$ 
would eventually tend to zero if the simulation could be carried out
for long enough.

The energy that the chromodyon loses through the slowing of the gauge
rotation must be carried away by radiation of the massless gauge
fields in the unbroken non-Abelian subgroup.  (Indeed, much of the
nonrotational contribution to $\eta$ at early times can presumably be
attributed to the creation of the radiation field.)  It is of interest
to know how this radiation depends on $\omega_0$.  We can use the
slope of the constant part in Fig.~\ref{fig:sim_results} to
approximate $d^2\theta/dt^2$ for $\omega_0 = 0.04ev$.  In
Table~\ref{radiationTable} we show the results for this as well as the
corresponding results for simulations with two other values of
$\omega_0$.  Recalling from Eq.~(\ref{energyOFrot}) that the energy
associated with the phase rotation is proportional to $\omega^2$, we
see that this data is consistent with
\begin{equation}
    {d E_{\rm ch} \over dt} \sim \omega_{\rm eff} {d\omega_{\rm eff} \over dt}
      \sim \omega_{\rm eff}^4  \, .
\end{equation}

\begin{center}\begin{table}[t]
\begin{tabular}{|c|c|c|} \hline
\hspace{0.5in}$\omega_0$\hspace{0.5in} 
&\hspace{0.5in} $d^2 \theta/ dt^2$\hspace{0.5in} 
&\hspace{0.35in} $(ev/\omega_0^3) \, d^2 \theta/ dt^2$ \hspace{0.35in} \\
\hline
$0.01\, ev$ & $8.0 \times 10^{-7} \,(ev)^2$ & 0.80  \\
$0.02\, ev$ & $6.2 \times 10^{-6}\,(ev)^2$ & 0.78  \\
$0.04\,ev$ & $4.6 \times 10^{-5}\,(ev)^2$ & 0.72  \\ 
\hline
\end{tabular}
\caption{Dependence of the deceleration of the initial global 
gauge rotation on the initial angular velocity $\omega_0$. }
\label{radiationTable}
\end{table}
\end{center}

We can understand this dependence by considering the energy flux carried by 
the radiation of the massless non-Abelian gauge fields in the unbroken 
subgroup.  This should be given by the analogue of the electromagnetic 
Poynting vector, 
\begin{equation}
T_i \, \sim \, \epsilon_{ijk} \mbox{Tr} \, \hat E_j  \hat B_k \, ,
\end{equation}
where the hats indicate the $O(1/r)$ radiation components of the
field strengths.   Because the initial chromodyon configurations that 
we constructed satisfied the static field equations to first order in $\omega_0$,
these radiation fields must be each at least second order in $\omega_0$,
so that $T_i \sim \omega^4_0$.

\section{Concluding remarks}
\label{conclusion}

Magnetic monopoles can be promoted to dyons by time-dependent
excitation of their U(1) global gauge zero modes.  In this paper we
have addressed the question of whether monopoles in theories with
non-Abelian unbroken symmetries can be promoted to chromodyons ---
monopoles with non-Abelian electric charge --- by a similar excitation
of their non-Abelian global gauge zero modes.  It has long been known
that the answer is negative if the magnetic charge has a non-Abelian
component, because there are then topological obstructions that
preclude the existence of a chromodyon.  However, there are also
monopoles with purely Abelian asymptotic magnetic charge, for which
there is no such obstruction, that could potentially have chromodyonic
counterparts.  We have examined one such case here, using a
constructive approach.  We started with a configuration with a
globally rotating non-Abelian phase, and thus a nonzero
chromo-electric charge, and then numerically evolved it to see whether 
it would settle down in a stable static solution.  In our simulations
we found instead that the effective rate of gauge rotation slows down,
so that the chromodyon continually loses energy and chromo-electric
charge.  Although we were not able to continue the simulation until
this charge was completely lost, every indication suggests that this
would be the final state of the system.

It is instructive to compare our results with those that would have
been obtained by applying our methods in the theory with SU(2) broken
to U(1), where we know that there is a dyon with Abelian electric
charge.  Because there is no analogue of the cloud radius zero mode,
with its associated complications, we can work in the BPS limit, where
analytic expressions are available.

Our approach would start with the static monopole solution
\begin{eqnarray}
    A_i^a &=& \epsilon_{aim} {\hat r_m} A(r) \, , 
      \cr \cr
    \Phi^a &=&  {\hat r_a} \, H(r) \, ,
\end{eqnarray}
with 
\begin{eqnarray}
   A(r) &=& {v \over \sinh evr} - {1\over er} \, , \cr
   H(r) &=& v \coth evr -{1\over er} \, .
\label{JZah}
\end{eqnarray}
Applying a global U(1) phase rotation and then gauge transforming
back to a static gauge would yield 
\begin{equation}
    (A_0^a)_{\rm static} = {\hat r_a} Q(r) \, , 
\end{equation}
where
\begin{equation}
     Q(r) = {\omega_0 \over ev} \left[v \coth evr -{1\over er} \right] \, .
\label{JZA0}
\end{equation}
From the $1/r$ term in this expression, we see that the asymptotic electric field
is 
\begin{equation}
   E_i^a =  \hat r_a \hat r_i\, {q_E \over r^2} \, ,
\end{equation}
where
\begin{equation}
    q_E = {1\over e} \left({\omega_0 \over ev}\right) \, .
\end{equation}
The energy of this configuration is 
\begin{equation}
    E = {4\pi v\over e} \left[1 + {e^2 q_E^2 \over 2} \right] \, ,
\end{equation}
where the first term represents the mass of the original monopole and the second
is the additional energy due to the phase rotation.

As in the chromodyon case, this initial configuration is only an approximate 
solution of the equations of motion.  The exact dyon solution with charge $q_E$ 
is given by~\cite{Prasad:1975kr,Bogomolny:1975de}
\begin{eqnarray}
    A(r) &=&  {v' \over \sinh ev'r} - {1\over er} \, , \cr \cr
    H(r)  &=& \cosh \gamma \left[v' \coth ev'r -{1\over er} \right] \, ,\cr \cr
    Q(r) &=& \sinh \gamma \left[v' \coth ev'r -{1\over er} \right] \, ,
\end{eqnarray}
where $\gamma$ is determined by the ratio of
electric and magnetic charges and is given by
\begin{equation}
    \sinh \gamma = e q_E 
\end{equation}
and $v' = v /\cosh \gamma$.
It has an energy 
\begin{equation}
    E = {4\pi v\over e} \sqrt{1 + (eq_E)^2 }
\end{equation}
and corresponds to phase rotating with an angular velocity
\begin{equation}
    \omega = ev' \sinh\gamma = { e^2vq_E \over \sqrt{ 1 + (eq_E)^2} } \, .
\end{equation}

Thus, our construction would start with a configuration that has a
core radius that is a factor $\sqrt{ 1 + (eq_E)^2}$ smaller than that
of the exact solution, and an angular velocity that is larger by the
same factor.  (The smaller core radius produces an decrease in the
phase rotation moment of inertia that exactly compensates for the
increase in angular velocity, thus yielding the same electric charge.)
The energy of this initial configuration exceeds that of the exact dyon
solution by an amount of order $q_E^4$.

It is easy to see what will happen if the initial configuration of
Eqs.~(\ref{JZah}) and (\ref{JZA0}) is allowed to evolve.  Because
radiation of the massive charged gauge field is energetically
suppressed, the electric charge of the dyonic configuration will be
conserved.  Hence, as the initial system relaxes it will tend toward
the static dyon solution of the same charge.  It will shed energy by
radiating massless photons, but the amount of this energy loss is
constrained by the fact that exact dyon mass places a lower bound on
the energy.  As the dyon radiates its phase rotation will slow down
and its core will expand.  For small electric charge (i.e., $eq_E \ll
1$), this slowing and expansion will both be small, and the system
will quickly approach its final state.  In particular, the slowing of
the phase rotation will be far less than that which we found in our
non-Abelian simulation.

Thus, our numerical simulations provide strong evidence against the
existence of static chromodyons in a theory with SO(5) broken to
SU(2)$\times$U(1).  Because it is hard to see how enlarging the
unbroken symmetry to a different non-Abelian group would stabilize the
chromodyon, we expect that similar results would hold for other
choices of gauge group and symmetry breaking.  Of course, numerical
simulations cannot provide a rigorous proof.  Even apart from issues
related to numerical accuracy, there is always the possibility that the
specific choice of initial configuration played a crucial role.  For
example, it is logically conceivable (although we think it quite
implausible) that there is some special choice or range of $\omega_0$
that would have led to a stable chromodyon.  Another possibility is
that there are chromodyon solutions, but that these exist only for
some minimum value of the chromo-electric charge.  In this case, the
solutions would not be continuously related to the purely magnetic
monopole, and so might not be found by our method.  Although we cannot
exclude this possibility, it seems to us to be rather unlikely.  Hence,
subject to these caveats, we conclude that static chromodyon solutions
do not exist.

\acknowledgments 
This work was supported in part by the U.S. Department of Energy.

\appendix

\section{Gauss's law equation with axial symmetry}
\label{apx:a0gausseqn}

As discussed in Sec.~\ref{sec:simInitialData}, when we gauge rotate
the spherically symmetric static non-BPS monopole solution, the
resulting Gauss's law equation has only axial symmetry and consists of
six coupled partial differential equations for the six coefficient
functions $u$, $w$, $b$, $Q$, $q$, and $t$ appearing in the $A_0$
ansatz of Eq.~(\ref{eqn:a0axialansatz}).  We write out the detailed
forms of these equations below.  Here,  $A$, $H$, $G$,
$K$, $F$, and $J$ are the functions, defined by Eq.~(\ref{ansatz}),
that specify the static non-BPS monopole solution and that are shown
in Fig.~\ref{fig:nonbpssol}.  To simplify the equations, we have 
set $e=1$ throughout; the explicit factors of $e$ can be recovered 
by simple dimensional analysis.

The two equations corresponding to first-sector components are
\begin{eqnarray}
\nonumber && \partial_{\rho \rho} u + {1 \over \rho} \partial_\rho u -
{u \over \rho^2} + \partial_{zz} u - 2 {\rho \over r} A \partial_z w +
2 {z \over r} A \partial_\rho w - 4 F \partial_z q + 4 F \partial_\rho
t \\ \nonumber && + 2 {\rho \over r} (\partial_r F) t - 2 {z \over r}
(\partial_r F) q - 6 {\rho \over r} A F t + 2 {\rho \over r} F G t -
{\rho \over r} K J t + {\rho \over r} H J t \\ \nonumber && + 2 {z
\over r} A F q - 2 {z \over r} F G q + {z \over r} K J q + 3 {z \over
r} H J q -2 {\rho^2 \over r^2} A^2 u - {z \rho \over r^2} A^2 w + {z
\rho \over r^2} H^2 w \\ && - {z^2 \over r^2} A^2 u - {z^2 \over r^2}
H^2 u - F^2 b + J^2 b - 3 F^2 u - J^2 u - 2 {1 \over r} A u = 0 \, ,
\\[12pt] \nonumber && \partial_{\rho\rho} w + {1 \over \rho}
\partial_\rho w + \partial_{zz} w + 4 F \partial_z t + 4 F
\partial_\rho q + 2 {\rho \over r} A \partial_z u - 2 {z \over r} A
\partial_\rho u + 2 {z \over r} (\partial_r F) t \\ \nonumber && + 2
{\rho \over r} (\partial_r F) q + 4 {1 \over \rho} F q - 2 {z \over
\rho} {1 \over r} A u - {\rho^2 \over r^2} A^2 w - {\rho^2 \over r^2}
H^2 w - 2 {z^2 \over r^2} A^2 w \\ \nonumber && - 2 {\rho \over r} A F
q + 2 {\rho \over r} F G q - {\rho \over r} K J q - 3 {\rho \over r} H
J q - 6 {z \over r} A F t + 2 {z \over r} F G t \\ && - {z \over r} K
J t + {z \over r} H J t -{\rho z \over r^2} A^2 u + {\rho z \over r^2}
H^2 u - F^2 Q + J^2 Q - 3 F^2 w - J^2 w = 0 \, .
\end{eqnarray}

The two second-sector equations can be obtained from these simply
by making the substitutions
\begin{eqnarray}
\nonumber u \rightarrow b \, , & & w \rightarrow Q \, , \\
A \rightarrow G \, , & & H \rightarrow K \, .
\end{eqnarray}

Finally, the third-sector equations are
\begin{eqnarray}
\nonumber && \partial_{\rho\rho} q + {1 \over \rho} \partial_\rho q -
{q \over \rho^2} + \partial_{zz} q + {\rho \over r} A \partial_z t -
{\rho \over r} G \partial_z t + F \partial_z b + F \partial_z u - {z
\over r} A \partial_\rho t + {z \over r} G \partial_\rho t \\
\nonumber && - F \partial_\rho Q - F \partial_\rho w - 2 {\rho \over
r} (\partial_r F) Q - {1 \over 2} {\rho \over r} (\partial_r F) w + {1
\over 2} {z \over r} (\partial_r F) b + {1 \over 2} {z \over r}
(\partial_r F) u \\ \nonumber && + {\rho \over r} \left[ {1 \over 2} A
F Q - {1 \over 2} F G Q + {1 \over 4} H J Q + {3 \over 4} J K Q - {1
\over 2} A F w + {1 \over 2} F G w - {3 \over 4} H J w - {1 \over 4} J
K w \right] \\ \nonumber && + {z \over r} \left[ - {1 \over 2} A F b +
{1 \over 2} F G b - {1 \over 4} H J b - {3 \over 4} J K b + {1 \over
2} A F u - {1 \over 2} F G u + {3 \over 4} H J u + {1 \over 4} J K u
\right] \\ && - {1 \over 2} A^2 q - {1 \over 2} G^2 q - {1 \over 4}
H^2 q - {1 \over 2} H K q - {1 \over 4} K^2 q - 4 F^2 q - 2 J^2 q - {1
\over r} A q - {1 \over r} G q = 0 \, ,
\end{eqnarray}
\begin{eqnarray}
\nonumber && \partial_{\rho\rho} t + {1 \over \rho} \partial_\rho t +
\partial_{zz} t + F \partial_z Q - F \partial_z w + F \partial_\rho b
- F \partial_\rho u - {\rho \over r} A \partial_z q + {\rho \over r} G
\partial_z q + {z \over r} A \partial_\rho q \\ \nonumber && - {z
\over r} G \partial_\rho q + {1 \over 2} {z \over r} (\partial_r F) Q
- {1 \over 2} {z \over r} (\partial_r F) w + {1 \over 2} {\rho \over
r} (\partial_r F) b - {1 \over 2} {\rho \over r} (\partial_r F) u \\
\nonumber && + {1 \over \rho} F b - {1 \over \rho} F u + {z \over
\rho} {1 \over r} A q - {z \over \rho} {1 \over r} G q \\ \nonumber &&
+ {\rho \over r} \left[ - {1 \over 2} A F b + {3 \over 2} F G b - {1
\over 4} H J b + {1 \over 4} J K b - {3 \over 2} A F u + {1 \over 2} F
G u + {1 \over 4} H J u - {1 \over 4} J K u \right] \\ \nonumber && +
{z \over r} \left[ - {1 \over 2} A F Q + {3 \over 2} F G Q - {1 \over
4} H J Q + {1 \over 4} J K Q - {3 \over 2} A F w + {1 \over 2} F G w +
{1 \over 4} H J w - {1 \over 4} J K w \right] \\ && - 6 F^2 t - {1
\over 2} A^2 t + A G t - {1 \over 2} G^2 t - {1 \over 4} H^2 t + {1
\over 2} H K t - {1 \over 4} K^2 t = 0 \, .
\end{eqnarray}

\section{Monitoring the Gauss's law constraint}
\label{apx:gausscheck}

Analytically, the Gauss's law constraint should remain satisfied
throughout the evolution if it is satisfied by the initial
data.  In our numerical calculation, we keep monitoring how well the
constraint is satisfied and use this as a way to check the
accuracy of our numerical methods.

The Gauss's law constraint is
\begin{equation}
D_j F^{j0} = ie[\Phi, D^0\Phi] \, .
\end{equation}
We can use Eq.~(\ref{eqn:evolveperturb}) to expand this into time-dependent
and time-independent parts.  Focusing on the time-dependent part, we have 
the requirement that 
\begin{eqnarray}
 0  &=& \partial_j {\tilde E_j} + ie[{\tilde A_j}, {\bar E_j}] +
ie[{\bar A_j}, {\tilde E_j}] + ie[{\tilde A_j}, {\tilde E_j}] 
  - \, ie[{\bar \Phi}, \partial_0 {\tilde \Phi}] - ie[{\tilde
\Phi}, \partial_0 {\tilde \Phi}] 
    \cr && - \, ie^2[{\bar \Phi}, i[A_0,
{\tilde \Phi}]] - ie^2[{\tilde \Phi}, i[A_0, {\bar \Phi}]] - ie^2[{\tilde
\Phi}, i[A_0, {\tilde \Phi}]] \, .
\label{eqn:gaussexpanded}
\end{eqnarray}

\begin{figure}[htp]
\begin{center}
\leavevmode \epsfxsize=4.6in \epsffile{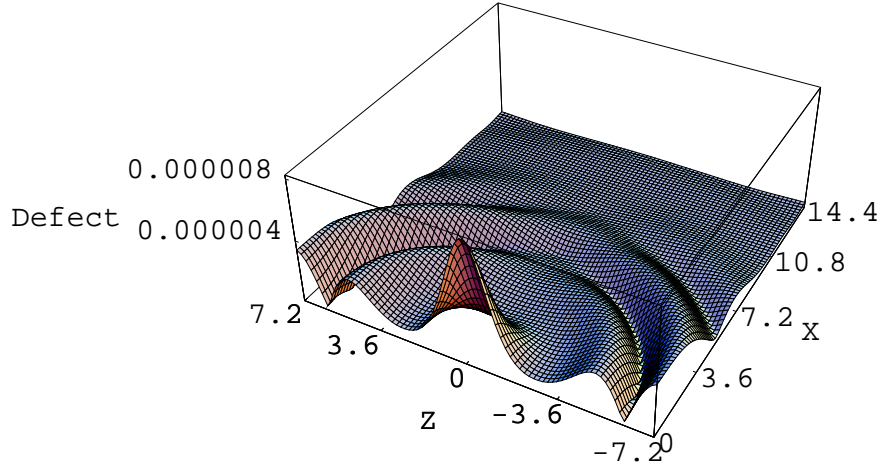}
\end{center}
\vspace{5pt}
\caption{The norm of the defect of the Gauss's law constraint at
$t=34.2(ev)^{-1}$. Since the system has an axial symmetry along the
$z$ axis, we plot the results on the $x$-$z$ plane. Distances are given in 
units of $1/ev$, and the defect is in units of $v$.}
\label{fig:gaussdefect}
\end{figure}

\begin{figure}[htp]
\begin{center}
\leavevmode
\epsfxsize=4.6in
 \epsffile{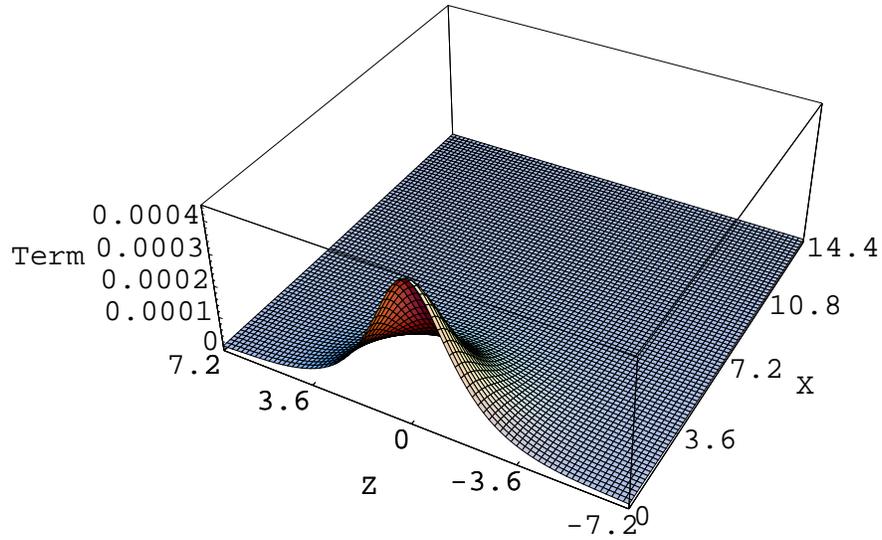}
\end{center}
\caption{The norm of the first term in Eq.~(\ref{eqn:gaussexpanded})
at $t=34.2(ev)^{-1}$. Since the system has an axial symmetry along the
$z$ axis, we plot the results on the $x$-$z$ plane. Distances are given in
units of $1/ev$, and the defect is in units of $v$.}
\label{fig:gaussterm}
\end{figure}

To demonstrate that the Gauss's law constraint is satisfied in our
numerical calculation, we calculate these nine terms using our
numerical results and then define their sum to be the defect.  Since
the defect is an so(5) element, we can define its norm to be the
square root of the trace of its square.  In
Figs.~\ref{fig:gaussdefect} and \ref{fig:gaussterm} we plot on the $x$-$z$
plane the
norm of the defect and, for comparison, the norm of the first term, both 
at $t=34.2 (ev)^{-1}$.  As we can see, the defect is about two
orders of magnitude smaller than the first term, indicating that the
Gauss's law constraint is satisfied very well.  We define the defect
level to be the ratio of the maximum norm of the defect and the
maximum norm among all terms. (Both maxima are found over the entire
computation domain.) For example, at $t=34.2(ev)^{-1}$, the maximum norm of the
defect is about $8\times 10^{-6}$, while the maximum norm of the first
term (which is larger than that of any of the other terms) is about
$4\times 10^{-4}$.  This give a defect level of about $0.02$.


\end{document}